\documentclass[11pt]{article}
\usepackage {amsfonts}
\usepackage {graphicx}
\usepackage {epsfig}
\usepackage{amssymb}
\usepackage{amsmath}
\usepackage {longtable}
\usepackage[english]{babel}
\begin{document}

\title{About possibility to search the electron EDM at the level $10^{-28} \div 10^{-30}$ e$\cdot$cm
and the constant of T-odd, P-odd scalar weak interaction of an
electron with a nucleus at the level $10^{-5} \div 10^{-7}$ in the
heavy atoms and ferroelectrics}

\author{V.G. Baryshevsky, \\ Research Institute for Nuclear Problems, Belarusian State
University,\\ 11 Bobruyskaya Str., Minsk 220050, Belarus,
\\ e-mail: bar@inp.minsk.by}

\maketitle

\begin{center}
\begin{abstract}
The T-odd phenomenon of induction of the magnetic field by a
static electric field provides to study the electron EDM and
constants of T-odd, P-odd interaction of an electron with a
nucleus.
Measurement of this magnetic field for ferroelectric materials
(like $PbTiO_3$) at the level $B \sim 3 \cdot 10^{-18}~G$ allows
to derive the electric dipole moment of an electron at the level
$d_e \sim 10^{-30}~e \cdot cm$ and the constant of T-odd scalar
weak interaction of an electron with a nucleus at the level
$k_1^{nuc} \sim 10^{-9}$.
The atomic magnetometry makes possible to measure fields $ \sim
10^{-13}~G/\sqrt{Hz}$ now. This means that for 10 days operation
one can expect to obtain $B$ at the level
\begin{equation}
B \sim 10^{-16}~G, \nonumber
\end{equation}
and, therefore, the  limits for $d_e$ in $PbTiO_3$ at the level
\begin{equation}
d_e \sim 10^{-28} ~{\text { and }} ~k_1^{nuc} \sim 10^{-7}.
\nonumber
\end{equation}
that makes the discussed method beneficial for measuring $d_e $
and $k_1^{nuc} $.

\end{abstract}
\end{center}

 \section{INTRODUCTION}

Study of time-reversal (T) symmetry violation (and CP (charge
conjugation, parity) symmetry violation {through} the CPT-theorem)
beyond the Standard model is of great importance for selecting the
fundamental interactions theory. This is the reason for focusing
efforts on the preparing and carrying out the dedicated
experiments. The searches of the EDM of an electron (atom,
molecule, nuclei) and T-odd weak interactions of electrons with
nuclei \cite{1}-\cite{4} are the most attractive.

In \cite{6,7} it was shown that measurement of the T-violating
polarizability $\beta^T$ (CP violating polarizability
$\beta^{CP}$) by the use of the phenomenon of the magnetic field
generation by a static electric field (and generation of the
electric field by a static magnetic field) could be a method for
such a search. Low temperatures are not required for these
measurements and this is the advantage of the above method.

\section{What values of $\beta^T$ could be measured in heavy atoms and
what values for the electron EDM and the constants of T-,P-odd
scalar weak interaction of an electron with a nucleon could be
derived from this measurement}

According to \cite{6,7} an external electric field acting on a
nonpolarized atom (molecule, nucleus) induces its magnetic dipole
moment $\vec{\mu}_E$:
\begin{equation}
\vec{\mu} (\vec E)=\beta_{s}^{T} \vec{E}, \label{8}
\end{equation}
where $\beta _{s}^{T}$ is the T-odd scalar polarizability of an
atom (molecule, nucleus).

 It
follows from (\ref{8}) that in a substance placed into an electric
field the magnetic field is induced \cite{6,7}:
\begin{equation}
\vec{B}_{E}^{ind}=f \rho \beta _{s}^{T} \vec{E}^{*}, \label{9}
\end{equation}
where $\rho$ is the substance density (i.e. the number of atoms
(molecules) per cm$^3$), $\vec{E}^{*}$ is the local field acting
on an atom (molecule) in the substance.
Parameter $f$ describes dependence of the macroscopic magnetic
field of the sample on the sample shape, for example:
\begin{equation}
{\text {for a sphere:}}~~f=\frac{8 \pi}{3}, ~ {\text {for a
cylinder:}}~~f=4 \pi. \nonumber
\end{equation}

If the substance consists of atoms of several types, then their
contributions to the induced field can be expressed as a sum of
contributions from different types of atoms:
\begin{equation}
\vec{B}_E=f \rho \sum\limits_{n}c_{n}\beta _{ns}^{T}
\vec{E}_n^{*}, \vec{E}_B=f \rho \sum\limits_{n}c_{n}\beta
_{ns}^{T} \vec{B}_{n}^{*}, \label{*}
\end{equation}
where $c_n$ is the concentration of atoms of the type $n$,
$\vec{E}_n^{*}$ and $\vec{B}_n^{*}$ are the local fields acting on
atoms of the type $n$.

According to \cite{6,7} the magnitude of the polarizability
$\beta_s^T$ is determined by both the electron EDM $d_e$ and T-odd
scalar weak interaction of an electron with a nucleus, which is
described by the constants $k_1^{p}$ and $k_1^{n}$ ($k_1^{p}$ is
the constant defining interaction with a proton, $k_1^{n}$ is that
for a neutron).
It is known \cite{8} that the expression for T-odd scalar weak
interaction of an electron with a nucleus includes constants
$k_1^p$ and $k_1^{n}$ as the sum $(Z k_1^{p}+ N k_1^{n})$, where
$Z$ and $N$ are the numbers of protons and neutrons in the
nucleus, respectively.
As a consequence, measurement of $\beta_s^T$ provides for getting
information about $d_e$ and the above sum \cite{6,7}.

Evaluations for $d_e$ and the sum $(Z k_1^{p} + N k_1^n)$ could be
derived from $\beta_s^T$.

Pioneering calculations of these contributions to the
polarizability $\beta_s^T$ for rare gas atoms were done in
\cite{9}.
It should be noted that in \cite{9} contribution of T-odd scalar
weak
 interaction was calculated as the function of  the constant
$k_1^{nuc}=\frac{1}{Z} (Z k_1^{p} + N k_1^n) =k_1^p+\frac{N}{Z}
k_1^n$.

 Let us evaluate now the least values of $\beta_s^T$ could be
 expected for measuring in the nearest future.

From (\ref{9}) it can be obtained
\begin{equation}
\beta_s^T=\frac{B_{E}}{f \rho E^{*}}
\end{equation}

 According to \cite{1} we can expect for magnetic induction measurement the sensitivity up to $\sim 3 \cdot
 10^{-18}G$.
 Such sensitivity of magnetic induction measurement provides for
 polarizability $\beta_s^T$ measurement the sensitivity:
\begin{eqnarray}
\begin{array}{c}
\beta_s^T=\frac{3 \cdot 10^{-18}}{f \rho E^{*}}
\end{array}
\nonumber
\end{eqnarray}

Let us consider first a non-ferroelectric material, for example:
liquid or solid $Xe$ in the highest available external field,
which does not exceed break-down voltage $E^{*} \sim 4 \cdot
10^{5}~V/cm=\frac{4}{3} \cdot 10^3$ CGSE. In this case (the
density of liquid and solid components $\rho \approx 2 \cdot
10^{22}~cm^{-3}$) $\beta_s^T$ can be evaluated as follows
\begin{eqnarray}
\begin{array}{r}
\beta_s^T \approx 10^{-44}~cm^3.
 \nonumber
\end{array}
\end{eqnarray}

Let us consider what limits for $d_e$ and  $k_1^{nuc}$ could be
got from $\beta_s^T$ measurement at such level.

According to \cite{9} contributions from $d_e$ and $k_1^{nuc}$
sharply depend on the nucleus charge $Z$ (as $Z^5$).
For example, from these calculations it follows that $\beta^T$ for
$Rn$ ($Z=86$) $20$ times exceeds $\beta^T$ for $Xe$ ($Z=54$) .
In particular, according to \cite{9} the contribution to the
polarizability $\beta_s^T$  caused by the electric dipole moment
of electron for $Xe$ atom ($Z=54$) can be expressed in atomic
units as follows:
\begin{equation}
\beta_{s~Xe}^T=1.5 \cdot 10^{-2}~d_e
\end{equation}
that can be rewritten in CGSE units as:
\begin{equation}
\beta_{s~Xe}^T=1.5 \cdot 10^{-2}~d_e [e \cdot cm] a_A^2 =1.5 \cdot 10^{-2}~d_e [e \cdot cm] (5 \cdot 10^{-9})^2,
\end{equation}
where $a_A$ is the atomic length unit.

Therefore, we obtain
\begin{equation}
d_e=\frac{\beta_{s~Xe}^T}{3.75 \cdot 10^{-19}} \approx 2.7 \cdot
10^{-26}
\end{equation}
(the similar evaluation $ d_e \sim 6 \cdot 10^{-26}$ is cited in
\cite{9}).

 At the same time for radon $Rn$ ($Z=80$) we
can get from the Table 1 in \cite{9}
\begin{equation}
\beta_{s~Rn}^T=d_e [e \cdot cm] 2.5 \cdot 10^{-17}
\end{equation}
i.e.
\begin{equation}
d_e = 4 \cdot 10^{-28}
\end{equation}
and this result raises hopes when keeping in mind that the
up-to-date limit for the electron EDM  $d_e(Tl)<1.6 \cdot
10^{-27}~e \cdot cm$ \cite{2}.

However, $Rn$ is nonstable and, therefore, it is not right for the
experiments. Though there are no calculations for other types of
atoms, the expectation to find the similar $Z^5$ dependence for
other atoms (for example, for $Pb~(Z=82)$ or $U~(Z=92)$) seems
also valid, because the lack (or excess) of number of electrons
$\Delta N_e$ with respect to such number in the closed shell is
much less than the total number of electrons in a heavy atom. This
is why the value of polarizability for $Pb~(Z=82)$ could be
supposed close to that for $Rn$ (while for $U$ it could be even
greater).

Thus, such evaluation for $Pb$ and $U$ polarizabilities gives hope
to get for the EDM measurement the limit $d_e \le 4 \cdot
10^{-28}~e \cdot cm$ (for the same external electric field $E \sim
4 \cdot 10^5~\frac{V}{cm}$). As $Pb$ and $U$ are metals, then, for
example, oxides should be used.

Let us consider now the limitations for the constant $k_1^{nuc}$
of electron-nucleus T-odd scalar weak interaction those could be
obtained from the above $\beta^T$ evaluation.

According to the Table 1 in \cite{9} the similar evaluations for
$Xe$ provide
\begin{equation}
{\text {in~ atomic ~units~}} \beta_S^T = 5.3 \cdot 10^{-15}~k_1
\nonumber
\end{equation}
\begin{equation}
{\text {in~ CGSE ~units~}} \beta_S^T = 5.3 \cdot
10^{-15}~k_1^{nuc} (a_A)^3 ~ (a_A=5 \cdot 10^{-9} ~cm) \nonumber
\end{equation}
Therefore,
\begin{equation}
k_1^{nuc} \approx 2 \cdot 10^{-5} \nonumber
\end{equation}
At the same time estimation for $Rn$ gives
\begin{equation}
k_1^{nuc} \approx 3 \cdot 10^{-7} . \nonumber
\end{equation}
Now the experimentally obtained limit for $Xe$ is $k_1^{nuc} \leq
10^{-4}$.
The experiment, which is planned with $Cs$ \cite{3}, could provide
for  $k_1^{nuc} \leq 5 \cdot 10^{-6}$ \cite{7}.

If again consider the evaluation made for $Rn$ close to those
could be obtained for $Pb$ and $U$ then the  ten times reduction
of limit for $k_1^{nuc}$ measurement could be expected in the
experiments with $Pb$ and $U$ oxides comparing with that is
planned with $Cs$ \cite{3}.

\section{Possible limit for $d_e$ and $k_1^{nuc}$ could be obtained in the ferroelectric $PbTiO_3$}

It is known that internal electric fields in ferroellectrics are
very large.
$PbTiO_3$ is a typical ferroelectric that has been considering as
a candidate for use in the EDM searches. According to evaluations
\cite{10} the field acting on the $Pb$ atom in $PbTiO_3$ can be
estimated as $E \approx 10^8~\frac{V}{cm}$. This means that
measuring the magnetic field induced by an external electric field
acting on a $PbTiO_3$-sample one can expect to measure $\beta_s^T$
at the level $\beta_s^T \sim 10^{-46}$ $cm^3$ (that is two orders
less than the above estimation $\beta_s^T \sim 10^{-44}$ $cm^3$).
Therefore the limits for $d_e$ and $k_1^{nuc}$ measurement are
\begin{equation}
d_e \sim 10^{-30} e \cdot cm {\text { and }} k_1^{nuc} \sim
10^{-9}. \nonumber
\end{equation}

\section{Conclusion}

Measurement of the magnetic field for ferroelectric materials
(like $PbTiO_3$) at the level $B \sim 3 \cdot 10^{-18}$ $G$ allows
to measure $\beta_s^T$  at the level $\beta_s^T \approx 10^{-46}
~cm^3$. This makes possible to derive the electric dipole moment
of an electron at the level $d_e \sim 10^{-30}~e \cdot cm$ and the
 constant of T-odd scalar weak interaction of an electron with a nucleus at the level
$k_1^{nuc} \sim 10^{-9}$.

Considering that the up-to-date limits are $d_e (Te)\leq
10^{-27}~e \cdot cm$ and $k_1^{nuc} \sim 10^{-4}$, the first step
goal could be posed as $d_e$ search at the level $10^{-28}~e \cdot
cm$ and, hence, reduce the requirements for $B$ measurement to $B
\sim 3 \cdot 10^{-16}~G$. An attempt to measure $k_1^{nuc} \sim
10^{-5}$ result in easier requirement for $B$ measurement $B \sim
3 \cdot 10^{-14}~G$. According to \cite{11} the atomic
magnetometry makes possible to measure fields $ \sim
10^{-13}~G/\sqrt{Hz}$ now. This means that for 10 days operation
one can expect to obtain $B$ at the level
\begin{equation}
B \sim 10^{-16}~G, \nonumber
\end{equation}
and, therefore, the  limits for $d_e$ in $PbTiO_3$ at the level
\begin{equation}
d_e \sim 10^{-28} {\text { and }} k_1^{nuc} \sim 10^{-7}.
\nonumber
\end{equation}
The above reasoning makes the proposed method beneficial for
measuring $d_e $ and $k_1^{nuc} $.


\end{document}